\documentstyle[preprint,aps,epsf,floats]{revtex}    
\def\bq{\begin{equation}}
\def\eq{\end{equation}}
\def\ba{\begin{eqnarray}}
\def\ea{\end{eqnarray}}

\begin{document}
\thispagestyle{empty}

\newcommand{\sla}[1]{/\!\!\!#1}

\renewcommand{\small}{\normalsize} 

\preprint{
\font\fortssbx=cmssbx10 scaled \magstep2
\hbox to \hsize{
\hskip.5in \raise.1in\hbox{\fortssbx University of Wisconsin - Madison}
\hfill\vtop{\hbox{\bf MADPH-98-1057}
            \hbox{\bf KEK-TH-589}
            \hbox{August 1998}} }
}

\title{\vspace{.5in}
Searching for $H \rightarrow \tau \tau$ in weak boson fusion
at the LHC
}
\author{D.~Rainwater \ and D.~Zeppenfeld\\[3mm]}
\address{
Department of Physics, University of Wisconsin, Madison, WI 53706
}
\author{K. Hagiwara}
\address{
Theory Group, KEK, Tsukuba, Ibaraki 305-0801, Japan
}
\maketitle
\begin{abstract}
Weak boson fusion is a copious source of intermediate mass Higgs bosons at
the LHC. The additional very energetic forward jets in these events provide
for powerful background suppression tools. We analyze the $H \to \tau\tau$
decay mode for the Standard Model Higgs boson. A parton level analysis of the 
dominant physics backgrounds (mainly $Z \to \tau\tau$ and Drell-Yan
production of $\tau$'s) and of reducible backgrounds (from $W+$~jet and
$b\bar{b}$ production in association with two jets and subsequent leptonic
decays) demonstrates that this channel allows the observation of
$H \to \tau\tau$ in a low background environment, yielding a significant
Higgs signal with an integrated luminosity of about 30~fb$^{-1}$. The weak
boson fusion process thus allows direct measurement of the $H\tau\tau$
coupling.
\end{abstract}
%
%

\newpage

%
%

\section{Introduction}\label{sec:one}

The search for the Higgs boson and, hence, for the origin of electroweak
symmetry breaking and fermion mass generation,
remains one of the premier tasks of present and future high energy physics
experiments. Fits to precision electroweak (EW) data have for
some time suggested a relatively small Higgs boson mass, of order
100~GeV~\cite{EWfits}. This is one of the reasons why the search for
an intermediate mass Higgs boson is particularly important~\cite{reviews}.
Beyond the reach of  LEP at CERN and of the Fermilab Tevatron, for masses
in the $110-150$~GeV range,
we show that observation of the $H \to \tau\tau$ decay channel at the
CERN Large Hadron Collider (LHC) is quite promising.  An advantage of the
$H \to \tau\tau$ channel, in particular compared to the dominant
$H \to b\bar{b}$ mode, is the lower background from QCD processes. The
$H \to \tau\tau$ channel thus offers the best prospects for a direct
measurement of the Higgs boson's couplings to fermions.

For the intermediate mass range, most of the literature has focussed on
Higgs production via gluon fusion~\cite{reviews}
and $t\bar{t}H$~\cite{ttH} or $WH(ZH)$~\cite{WH} associated production.
Cross sections for Higgs production at the LHC are well-known~\cite{reviews}, 
and while production via gluon fusion has the largest cross section by almost 
one order of magnitude, there are substantial QCD backgrounds but
few handles to distinguish them from the signal.  Essentially, only the decay 
products' transverse momentum and the resonance in their
invariant mass distribution can be used. The second largest production
cross section for the standard model (SM) Higgs boson is predicted for
weak-boson fusion (WBF), $qq \to qqVV \to qqH$. WBF events contain
additional information in the observable quark jets. Techniques like
forward jet tagging~\cite{Cahn,BCHP,DGOV} can then be exploited to reduce
the backgrounds.

Another feature of the WBF signal is the lack of color
exchange between the initial-state quarks. Color coherence between initial-
and final-state gluon bremsstrahlung leads to suppressed hadron production in 
the central region, between the two tagging-jet candidates of the
signal~\cite{bjgap}. This is in contrast to most background processes, which
typically involve color exchange in the $t$-channel and thus lead to enhanced 
hadronic activity in the central region. We exploit these features, via
a veto on additional soft jet activity in the central
region~\cite{bpz_minijet}.

While some attention has been given to $A/H \to \tau\tau$ searches at the
LHC~\cite{Cavalli,RW,CMS-ATLAS}
in the framework of the MSSM, where the increased
couplings of A/H to $\tau$ predicted for $\tan\beta \gg 1$ lead to
higher production rates, conventional wisdom
says that the chance of seeing the SM Higgs via this decay mode is nil,
and it has heretofore been ignored in the literature.
Thus, we provide a first analysis of intermediate-mass SM $H \to \tau\tau$ at 
the LHC (and of the main physics and reducible backgrounds) which 
demonstrates the feasibility of Higgs boson detection in this channel,
with modest luminosity. $H \to \tau\tau$ event characteristics are analyzed
for one $\tau$ decaying leptonically and the other decaying hadronically,
because of the high trigger efficiency and good branching ratio of this
mode; Ref.~\cite{Cavalli} found the dual leptonic decay
mode to be considerably more difficult due to higher backgrounds.

Our analysis is a parton-level Monte Carlo study, using full tree-level
matrix elements for the weak boson fusion Higgs signal and the various
backgrounds. In Section~\ref{sec:two} we describe our calculational
tools, the methods employed in the simulation of the various processes,
and important parameters. Extra minijet
activity is simulated by adding the emission of one extra parton to the
basic signal and background processes. Generically we call the basic signal
process (with its two forward tagging jets) and the corresponding background
calculations ``2-jet'' programs, and refer to the simulations with one extra
parton as ``3-jet'' programs. In
Sections~\ref{sec:three}~\&~\ref{sec:four}, using the 2-jet programs for
physics and reducible backgrounds, respectively, we demonstrate forward jet
tagging and $\tau$ identification and reconstruction criteria which
yield an $\approx$2/1 signal-to-background (S/B) ratio.
Both the $Wj+jj$ and $b\bar{b}jj$ reducible backgrounds intrinsically are
much larger than the $Z \to \tau\tau$ and
Drell-Yan $\tau$-pair production backgrounds. We
explain and emphasize the cuts crucial to reducing these backgrounds to a
manageable level.

In Section~\ref{sec:five} we analyze the different minijet patterns in signal 
and background, using the truncated shower approximation (TSA)~\cite{TSA} to
regulate the cross sections. By exploiting the two most important
characteristics of the extra radiation, its angular distribution and its
hardness, the QCD
backgrounds can be suppressed substantially by a veto on extra central jet
emission. Within the TSA,
probabilities are estimated for vetoing signal and background events,
and are combined with the production cross sections of the previous section
to predict signal and background rates in Table~\ref{table4}. These rates
demonstrate the possibility to extract a very low background $H\to\tau\tau$
signal at the LHC.

Our signal selection is not necessarily optimized yet. Additional observables
are available to distinguish the signal from background. The final discussion 
in Section~\ref{sec:six} includes a survey of distributions which can be
used, e.g. in neural-net algorithms, to further improve the signal 
significance.

\section{Calculational Tools }\label{sec:two}

We simulate $pp$ collisions at the CERN LHC, $\protect\sqrt{s} = 14$~TeV.
All signal and background cross sections are determined in terms of full
tree level matrix elements for the contributing subprocesses and will be
discussed in more detail below.

For all our numerical results we have chosen ${\rm sin}^2\theta_W = 0.2315$,
$M_Z = 91.19$~GeV, and $G_F = 1.16639\cdot 10^{-5}\;{\rm GeV}^{-2}$, which
translates into $M_W = 79.97$~GeV and $\alpha(M_Z) = 128.93$ when using
the tree-level relations between these input parameters. This value of $M_W$
is somewhat lower than the current world average of $\approx 80.35$ GeV.
However, this difference has negligible effects on all cross sections, e.g.
the $qq\to qqH$ signal cross section varies by about $0.5\%$ for these two
$W$ mass values. The tree level relations between the input parameters are
kept in order to guarantee electroweak gauge invariance of all amplitudes.
For all QCD effects, the running of the strong-coupling constant is
evaluated at one-loop order, with $\alpha_s(M_Z) = 0.118$. We employ
CTEQ4L parton distribution functions~\cite{CTEQ4_pdf} throughout. Unless
otherwise noted the factorization scale is chosen as $\mu_f =$ min($p_T$)
of the defined jets.

\subsection{The $qq\to qqH(g)$ signal process}

The signal can be described, at lowest order, by two single-Feynman-diagram
processes, $qq \to qq(WW,ZZ) \to qqH$, i.e. $WW$ and $ZZ$ fusion where the
weak bosons are emitted from the incoming quarks~\cite{qqHorig}.
From a previous study of $H\to\gamma\gamma$ decays in weak boson
fusion~\cite{RZ_gamgam} we know several features of the signal, which we can
directly exploit here: the centrally produced Higgs boson tends to yield
central decay products (in this case $\tau^+\tau^-$), and the two quarks
enter the detector at large rapidity compared to the $\tau$'s
and with transverse momenta in the 20 to 80 GeV range, thus leading to two
observable forward tagging jets.

For the study of a central jet veto, the emission of at least one extra
parton must be simulated. This is achieved by calculating the cross sections
for the process $qq\to qqHg$, i.e. weak boson fusion with radiation of an
additional gluon, and all crossing related process. These include
\bq
q\bar{q} \to q\bar{q}Hg \, ,\qquad \;
\bar q\bar q \to \bar q\bar qHg \, , \qquad
qg \to qq\bar{q}H \, , \qquad
\bar q g \to \bar qq\bar{q}H \, ,
\label{procsig}
\eq
and can be found in Ref.~\cite{qqHjorig}.
For this case with three final state partons, the factorization scale
is chosen as $\mu_f =$ min($p_T$) of the tagging jets
and the renormalization scale $\mu_r$ is set to the transverse momentum
of the non-tagging parton (minijet).
Different scale choices or different input parameters will, of
course, affect our numerical results. Variation of the factorization
scale by a factor of two changes the 2-jet cross section
in the last column of Table~\ref{table1} by $\leq \pm 10\%$.

In the following we only consider $\tau$-pair decays with one $\tau$ decaying
leptonically, $\tau \to e\nu_e\nu_\tau,\; \mu\nu_\mu\nu_\tau$, and the other
decaying hadronically, $\tau^\pm\to h^\pm X$, with a combined branching
fraction of $45\%$. Our analysis critically employs
transverse momentum cuts on the charged $\tau$-decay products and, hence,
some care must be taken to ensure realistic momentum distributions.

Because of its small mass, we simulate the $\tau$ decays in the collinear
approximation. The momentum fraction $z$ of the charged decay lepton in
$\tau^\pm \to \ell^\pm\nu_\ell\nu_\tau$ is generated according to the
decay distribution
\bq
{1\over \Gamma_\ell} {d\Gamma_\ell\over dz} =
{1\over 3} (1-z) \left[ (5+5z -4z^2) + \chi_\tau(1+z-8z^2) \right]\; .
\eq
Here $\chi_\tau$ denotes the chirality of the decaying $\tau$ (which,
for a negative helicity $\tau^-$ or positive helicity $\tau^+$, is given
by $\chi_\tau=-1$  in the collinear limit). Similarly the
pion spectrum for $\tau^\pm \to \pi^\pm\nu_\tau$ decays is given by
\bq
{1\over \Gamma_\pi } {d\Gamma_\pi\over dz} \simeq 1 + \chi_\tau(2z-1)\; .
\eq
Decay distributions for $\tau\to\rho\nu_\tau$ and $\tau\to a_1\nu_\tau$
are taken from Ref.~\cite{HMZ}. We add the decay distributions from the
various hadronic decay modes according to their branching ratios. The
vector meson decays are simulated in the narrow width approximation, which
is adequate for our purposes.
The decay of the Higgs scalar produces $\tau$'s of opposite chirality,
$\chi_{\tau^+}=-\chi_{\tau^-}$ and this anti-correlation of the $\tau^\pm$
polarizations is taken into account.

Positive identification of the hadronic $\tau^\pm\to h^\pm X$ decay requires
severe cuts on the charged hadron isolation. Possible strategies have been
analyzed by Cavalli {\it et al.}~\cite{Cavalli} and we base our simulations
on their result. Considering hadronic
jets of $E_T>40$~GeV in the ATLAS detector, they find non-tau rejection
factors of 400 or more (see below) while retaining true hadronic $\tau$ decays
with an identification efficiency
\bq\label{eq:epstauh}
\epsilon_\tau(\tau\to\,\nu+{\rm hadrons}) = 0.26\; .
\eq
This estimate includes the requirement of seeing a single charged hadron
track, of $p_T>2$~GeV, pointing in the $\tau$ direction, and thus effectively
singles out 1-prong $\tau$ decays. Accordingly, only the 1-prong hadronic
branching ratios are considered in our mixture
of $\pi$, $\rho$ and $a_1$ modes.
Since the overall efficiency includes 3-prong events, which have negligible
acceptance, the effective efficiency for 1-prong events is larger and taken
as 0.34 in the following, which reproduces the overall efficiency of
Eq.~(\ref{eq:epstauh}).

\subsection{The QCD $\tau^{+}\tau^{-}+jj(j)$ physics background}

Given the H decay signature, the main physics background to our signal
$\tau^+ \tau^- jj$ events arises from real emission QCD corrections to
the Drell-Yan process $q\bar{q} \to (Z,\gamma) \to \tau^+\tau^-$.
For $\tau^+ \tau^- jj$ events these background processes include~\cite{Kst}
\bq\label{procback}
q g \to q g \tau^+ \tau^- \, , \qquad  q q' \to q q' \tau^+ \tau^- \, ,
\eq
which are dominated by $t$-channel gluon exchange, and all crossing
related processes, such as
\bq
q \bar{q} \to g g \tau^+ \tau^- \, , \qquad g g \to q \bar{q} \tau^+ \tau^- \;.
\eq
All interference effects between virtual photon and $Z$-exchange are included.
The $Z$ component dominates, however, and we call these processes
collectively the ``QCD $Zjj$'' background. The cross sections for the
corresponding $Z+3$-jet processes, which we need for our modeling of minijet
activity in the QCD $Zjj$ background, have been calculated in
Refs.~\cite{HZ,BHOZ,BG}. Similar to the treatment of the signal processes, we 
use a parton-level Monte-Carlo program based on the work of Ref.~\cite{BHOZ}
to model the QCD $Zjj$ and $Zjjj$ backgrounds.

The factorization scale is chosen as for the Higgs boson signal.
With $n=2$ and $n=3$ colored partons in the final state,
the overall strong-coupling constant factors are taken as
$(\alpha_s)^n = \prod_{i=1}^n \alpha_s(p_{T_i})$, {\em i.e.} the transverse
momentum of each additional parton is taken as the relevant scale for its
production, irrespective of the hardness of the underlying scattering event.
This procedure guarantees that the same $\alpha_s^2$ factors are used for the 
hard part of a $Zjj$ event, independent of the number of additional minijets, 
and at the same time the small scales relevant for soft-gluon emission are
implemented.

The momentum distributions for the $\tau$ decay products are generated as
for the Higgs boson signal. Because of the (axial)-vector coupling of the
virtual $Z,\gamma$ to $\tau$'s, the produced $\tau^+$ and $\tau^-$ have the
same chirality. This correlation of the $\tau$ polarizations is taken into
account by calculating individual
helicity amplitudes and folding the corresponding cross sections with the
appropriate $\tau^+$ and $\tau^-$ decay distributions, i.e. the full $\tau$
polarization information is retained in the energy distribution of the $\tau$
decay products.

\subsection{The EW $\tau^{+}\tau^{-}+jj(j)$ physics background}

These backgrounds arise from $Z$ and $\gamma$ bremsstrahlung in
quark--(anti)quark scattering via $t$-channel electroweak boson exchange,
with subsequent decay $Z,\gamma\to \tau^+\tau^-$:
\bq
qq' \to qq' \tau^+\tau^-
\label{eq:qQqQZ}
\eq
Naively, this EW background may be thought of as suppressed compared to
the analogous QCD process in Eq.~(\ref{procback}).
However, the EW background includes electroweak boson fusion,
$VV \to \tau^+\tau^-$, either via $t$-channel $\tau/\nu$-exchange or via
$s$-channel $\gamma/Z$-exchange,
and the latter has a momentum and color structure which is identical
to the signal and cannot easily be suppressed via cuts.

We use the results of Ref.~\cite{CZ_gap} for our calculation which ignore
$s$-channel EW boson exchange contributing to $q\bar{q}$ production,
and Pauli interference of identical quarks.
When requiring a large rapidity separation between the two quark jets
(tagging jets) the resulting large dijet invariant mass severely suppresses
any $s$-channel processes which might give rise to the dijet pair, and the
very different phase space regions of the two scattered quarks make Pauli
interference effects small.
All charged-current (CC) and neutral-current (NC) subprocesses are included.
The CC process dominates over NC exchange, however, mainly because of
the larger coupling of the quarks to the $W$ as compared to photon
and $Z$ interactions.
As in the QCD $Zjj$ case, the $Z$-pole dominates the $\tau^+\tau^-$
invariant mass distribution, so we will refer to these EW processes as
the ``EW $Zjj$'' background.

The $\tau$ decay distributions are generated in the same way as described
above for the Higgs signal. Since the programs of Ref.~\cite{CZ_gap} generate
polarization averaged $\tau^+\tau^-$ cross sections, we have to assume
unpolarized $\tau$'s. However, as for the QCD $Zjj$ background, the
$\tau^+\tau^-$ pair arises from virtual vector boson decay, resulting in
a $\tau^+$ and $\tau^-$ of the same chirality.  This correlation of the
$\tau$ polarizations is taken into account.

In order to determine the minijet activity in the EW $Zjj$ background
we need to evaluate the ${\cal O}(\alpha_s)$ real parton emission
corrections. The corresponding ${\cal O}(\alpha^4\alpha_s)$ diagrams for
\bq
qq'\to qq'g\;\tau^+\tau^- \; ,
\label{eq:qQqQgtautau}
\eq
and all crossing related subprocesses, have first been calculated
in Ref.~\cite{RSZ_vnj}.
Production of the $\tau$-pair via $Z$ and $\gamma$ exchange is considered.
The factorization and renormalization scales are chosen to be the same as for 
the $Hjj$ signal, as this is also a hard EW process.

We have previously examined other scale choices for the $Z$
backgrounds~\cite{RSZ_vnj}, and found small uncertainties
($\approx \pm 15 \%$) for the EW component, while variations for the
QCD component reach a factor 1.5. We thus expect the signal and EW
$Zjj$ background cross sections to be fairly well determined at leading
order, while the much larger theoretical uncertainty for the QCD $Zjj$
background emphasizes the need for experimental input.

\subsection{The QCD $Wj+jj(j)$ reducible background}
\label{sec:twoD}

Reducible backgrounds to $\tau^+\tau^-\to \ell^\pm h^\mp\sla p_T$ events
can arise from any process with a hard, isolated lepton, missing $p_T$,
and an additional narrow jet in the final state which can be mistaken as a
hadronically decaying $\tau$. A primary reducible background thus arises from
leptonic $W$ decays in $Wj$ events, where additional QCD radiation supplies
the two tagging jet candidates. At lowest order we need to consider
$Wj+jj$ production as the hard process, which is very similar to the
simulation of the QCD $Zjjj$ background discussed before,
with the bremsstrahlung $Z$ replaced by a $W$. $W\to e\nu_e,\;\mu\nu_\mu$
decays only are considered and are treated as a fake $\tau$ decaying
leptonically. Real leptonic $\tau$ decays from
$W\to\tau\nu_\tau\to \ell\nu_\ell\nu_\tau$ are relatively suppressed by
the $\tau$ leptonic branching ratio of $35\%$ and the severity of the
transverse momentum cuts on the softer charged lepton spectrum. They will
be ignored in the following.

Two of the jets in $Wj+jj$ events are identified as tagging jets, and
fluctuations of the third
into a narrow jet are considered, resembling a hadronically-decaying $\tau$.
In Ref.~\cite{Cavalli} the probability for misidentifying a gluon or 
light-quark jet as a hadronic $\tau$ decay was estimated as 
\bq\label{eq:epsjtau}
\epsilon_\tau({\rm jet}\to\,''\nu+{\rm hadrons}'') = 0.0025 \; ,
\eq
and we assign this probability to each of the final state jets. 
In each event one of the hard partons is randomly assigned to be the $\tau$. 
To mimic the signal, this jet and the identified charged lepton must be of
opposite charge. Thus, we reduce the $Wj+jj$ background by an additional 
factor of two to simulate the opposite charge requirement for 
the single track allowed in the $\tau$-like jet. 
As the $Wj+jj$ events are a QCD background, we use the same 
factorization and renormalization scales as for the QCD $Zjj$ case.

To simulate additional minijet emission, we need to add one more parton to
the final state. The code for $W+4j$ matrix elements has been available since 
the work of Berends et al.~\cite{tausk}. Here we use the program developed
in Ref.~\cite{W4j}, which was generated via MadGraph~\cite{Madgraph}. Since
$W+4j$ production produces a six-particle final state, with up to 516 graphs
for the most complicated processes, it takes considerable CPU time to obtain
good statistics. We modified the MadGraph code to
do random helicity summation, speeding up the calculation by approximately a
factor of 3 for a given statistical error in the final cross section.
As before, $\alpha_s$ is taken as the geometric mean of
$\alpha_s(p_T)$ factors for each of the partons, including the parton
which fakes the hadronic $\tau$ decay.

\subsection{The QCD $b\bar{b}jj$ reducible background}
\label{sec:twoE}

The semileptonic decay of $b$-quarks provides another source of leptons
and neutrinos which can be misidentified as tau decays. Even though $b$-quark
decays are unlikely to lead to isolated charged leptons and very narrow
tau-like jets in a single event, the sheer number of $b\bar b$ pairs produced
at the LHC makes them potentially dangerous. Indeed, the analysis of
Ref.~\cite{Cavalli} found that $b\bar b$ pairs lead to a reducible
$\tau^+\tau^-$ background which is similar in size to $Wj$ production.
We therefore study $b\bar{b}jj$ production as our second reducible
background and neglect any other sources like $t\bar t$ events which
were shown to give substantially smaller backgrounds to
$\tau^+\tau^-$-pairs in Ref.~\cite{Cavalli}.

We only consider $b$-production events where both $b$-quarks have large
transverse momentum. In addition, two forward tagging jets will be
required as part of the signal event selection. The relevant leading
order process therefore is the
production of $b\bar{b}$ pairs in association with two jets, which
includes the subprocesses
\ba
    gg          & \rightarrow & b\bar{b} gg  \nonumber \\
    qg          & \rightarrow & b\bar{b} qg   \\
    q_{1} q_{2} & \rightarrow & b\bar{b} q_{1} q_{2} \,. \nonumber
\ea
The exact matrix elements for the ${\cal O} (\alpha_{s}^{4})$ processes
are evaluated, including all the crossing related subprocesses, and retaining
a finite $b$-quark mass~\cite{Stange}.
The Pauli interference terms between identical quark flavors in the process
$q_{1}q_{2}\rightarrow b\bar{b}q_{1}q_{2}$ are neglected, with little effect
in the overall cross section, due to the large differences in the rapidity
of the final state partons.
The factorization scale is chosen as the smallest transverse energy
of the final state partons before the $b$-quark decay.
The strong coupling constant $\alpha_{s}$
is evaluated at the corresponding transverse energy of the final
state partons, i.e.,
$\alpha_{s}^{4} = \alpha_{s}(E_{T}(b)) \alpha_{s}(E_{T}(\bar{b}))
                 \alpha_{s}(p_{T,{\rm jet}_{1}})
                 \alpha_{s}(p_{T,{\rm jet}_{2}})$.

The semileptonic decay $b\to\ell\nu c$ of one of the $b$-quarks is simulated
by multiplying the $b\bar{b}jj$ cross section by a branching ratio factor of
0.395 (corresponding to at least one semileptonic $b$-decay to occur) and by
implementing a three-body phase space distribution for the decay momenta.
This part of the simulation is performed in order to estimate the
effects of the lepton isolation cuts on the transverse momentum distributions
of the $b$-decay leptons. Since these are kinematic effects we use the lightest
meson masses in the simulation and set $m_b = 5.28$~GeV and $m_c=1.87$~GeV.
In Ref.~\cite{Cavalli} a factor 100 reduction of the $b\bar{b}$ background
was found as a result of lepton isolation, requiring $E_T<5$~GeV in a cone
of radius 0.6 around the charged lepton. In our simulation, after energy
smearing of the charm quark jet (see below), we find a reduction factor of 52
due to lepton isolation with a cone of radius 0.7. However, our simulation
does not include parton showers or hadronization of the $b$-quark, effectively
replacing the $b$-quark fragmentation function by a delta-function at one, and
thus underestimates the effect of lepton isolation cuts on the $b$-quark
background. To compensate for this, we multiply our $b\bar{b}jj$ rates by
another factor 0.52, thus effectively implementing the factor 100
suppression found by Cavalli {\it et al.}~\cite{Cavalli}.

In addition to an isolated lepton, the $b\bar{b}jj$ events must produce a
narrow jet which is consistent with a hadronic $\tau$ decay, and has charge
opposite the identified charged lepton. This may either be one of the light 
quark or gluon jets, for which the misidentification probability of $0.25\%$ 
of Eq.~(\ref{eq:epsjtau}) will be used, or it may be the $b$-quark jet. 
In Ref.~\cite{Cavalli} the probability for misidentifying
a $b$-quark jet as a hadronic $\tau$ decay was estimated as
\bq\label{eq:epsbtau-cavalli}
\epsilon_\tau(b\to\,''\nu+{\rm hadrons}'') \approx 0.0005 \; .
\eq
However, due to limited Monte Carlo statistics, this number was based on a
single surviving event only. Since we are really interested in an upper
bound on the $b\bar{b}jj$ background we follow the ATLAS
proposal~\cite{CMS-ATLAS} instead, and use the upper bound,
\bq\label{eq:epsbtau-ATLAS}
\epsilon_\tau(b\to\,''\nu+{\rm hadrons}'') < 0.0015 \; ,
\eq
for our analysis. Thus, all our $b\bar{b}jj$ cross sections, after $\tau$
identification, should be considered conservative estimates. A more precise
analysis of $b\to \tau$ misidentification probabilities in the LHC detectors 
is clearly needed, which is beyond the scope of the present work. Finally,
an additional overall factor of two reduction is applied, as in the $Wj+jj$
case, for the lepton-jet opposite charge requirement.

The purpose of our $b$-analysis is to verify that $b$ semileptonic decays do
not overwhelm the signal. The above procedures are adequate for this purpose,
since we obtain final $b\bar{b}jj$ backgrounds (in Table~\ref{table4}) which
are 20 to 40 times smaller than the signal.
We do not calculate additional $b$ quark backgrounds arising from intrinsic
$b$ contributions (processes like $gb\to bggg$). The matrix elements for these
processes are of the same order ($\alpha_s^4$) as for the $b\bar{b}jj$
subprocesses discussed above, but they are suppressed in addition by the
small $b$-quark density in the proton. Also, we do not simulate additional
soft gluon emission for the $b\bar{b}jj$ background. This would require
$b\bar{b}+3$~jet matrix elements which are not yet available.
Rather, in Section~\ref{sec:five}, we assume the probability for
extra minijet emission to be
the same as for the other reducible QCD background, $Wj+jj$ production.

\subsection{Detector resolution}

The QCD processes discussed above lead to steeply falling jet transverse
momentum distributions. As a result, finite detector resolution can have a
sizable effect on cross sections. Resolution effects are particularly
pronounced for the $b\bar{b}jj$ background, where a higher momentum charm
quark (from $b\to c\ell\nu$ decay) can fluctuate below the $E_T(c)<5$~GeV
isolation requirement of the charged lepton.

These resolution effects are taken into account via Gaussian smearing of
the energies of jets and $b$ and $\tau$ decay products.  Following ATLAS
expectations~\cite{CMS-ATLAS} we use resolutions
\bq
{\triangle{E} \over E} =
{5.2 \over E} \oplus {0.16 \over {\sqrt E}} \oplus .009 \, ,
\eq
for jets (with individual terms added in quadrature), while for charged
leptons we use
\bq
{\triangle{E} \over E} = 2\% \, .
\eq

In addition, finite detector resolution leads to fake missing transverse
momentum in events with hard jets. An ATLAS analysis~\cite{Cavalli} showed
that these effects are well parameterized by a Gaussian distribution of
the components of the fake missing transverse momentum vector,
$\vec\sla p_T$, with resolution
\bq
\sigma(\sla p_x,\sla p_y) = 0.46 \cdot \sqrt{\sum{E_{T,had}}} \, ,
\eq
for each component.
In our calculations, these fake missing transverse momentum vectors are added
linearly to the neutrino momenta.

\section{Higgs signal and real $\tau^{+}\tau^{-}$ backgrounds}
\label{sec:three}

The $qq\to qqH,\;H\to\tau\tau$ signal is characterized by two forward
jets and the $\tau$ decay products. Before discussing background levels
and further details like minijet radiation patterns, we need to identify
the search region for these hard $Hjj$ events. Prior to $\tau$ identification,
the task is identical to the Higgs search in $qq\to qqH,\;H\to\gamma\gamma$
which was considered previously~\cite{RZ_gamgam}. We can thus adopt the
strategy of this earlier analysis and start out by discussing three levels
of cuts on the $qq\to qqH,\;H\to\tau\tau$ signal, before considering $\tau$
decay and $\tau$ identification. This procedure makes explicit the source of
the major signal reduction factors which we will encounter.

The basic acceptance requirements must ensure that the two jets and
two $\tau$'s are observed inside the detector (within the hadronic and
electromagnetic calorimeters, respectively), and are well-separated from each
other:
\ba
& p_{T_{j(1,2)}} \geq 40, 20~{\rm GeV} \, ,\qquad |\eta_j| \leq 5.0 \, ,\qquad 
\triangle R_{jj} \geq 0.7 \, , & \nonumber\\
& |\eta_{\tau}| \leq 2.5 \, , \qquad \triangle R_{j\tau} \geq 0.7 \, , \qquad
\triangle R_{\tau\tau} \geq 0.7 \, . &
\label{eq:basic}
\ea
Slightly more than half of all signal events pass these basic cuts.
The staggered $p_{T_j}$ cuts anticipate the steeply falling transverse
momentum distributions of both jets for the QCD backgrounds, which are
dominated by bremsstrahlung gluons. In contrast, for the $Hjj$ signal,
the $p_T$ scale is set by the mass of the exchanged
weak bosons and most of the tagging jets survive these cuts.

Another feature of the irreducible QCD background is the generally higher
rapidity of the $\tau$'s as compared to the Higgs signal: $Z$ and $\gamma$
bremsstrahlung occur at small angles with respect to the parent quarks,
producing $\tau$'s forward of the jets. Thus, at the second level of cuts
we require both $\tau$'s to lie between the jets with a
separation in pseudorapidity $\triangle \eta_{j,\tau} > 0.7$, and the jets to 
occupy opposite hemispheres:
\bq
\label{eq:taucen}
\eta_{j,min} + 0.7 < \eta_{\tau_{1,2}} < \eta_{j,max} - 0.7 \, , \qquad
\eta_{j_1} \cdot \eta_{j_2} < 0
\eq
At the third level of cuts, which is also the starting point for our
consideration of the various backgrounds, a wide separation in
pseudorapidity is required between the two forward tagging jets,
\bq
\label{eq:gap}
\triangle \eta_{tags} = |\eta_{j_1}-\eta_{j_2}| \geq 4.4 \, ,
\eq
leaving a gap of at least 3 units of pseudorapidity
in which the $\tau$'s can be observed. This
technique to separate weak boson scattering from various backgrounds is
well-established~\cite{Cahn,BCHP,DGOV,bpz_minijet,RZ_gamgam}, 
in particular for heavy Higgs boson searches.
Table~\ref{table1} shows the effect of these cuts on the signal for a SM Higgs
boson of mass $m_H = 120$~GeV. Overall, about $25\%$ of all $H\to\tau\tau$
events generated in weak boson fusion are accepted by the cuts of
Eqs.~(\ref{eq:basic}-\ref{eq:gap}).

\begin{table}
\caption{Signal $H \to \tau\tau$ branching ratio times cross sections for
$m_H = 120$~GeV $Hjj$ events in $pp$ collisions at $\protect\sqrt{s}=14$~TeV. 
Results are given for successive cuts of Eqs.~(\ref{eq:basic}-\ref{eq:gap}).}
\label{table1}
\begin{tabular}{lccc}
\phantom{generic} &
           Eq.~\ref{eq:basic} & + Eq.~\ref{eq:taucen} & + Eq.~\ref{eq:gap} \\
\hline
B($H \to \tau^+\tau^-$)$\cdot \sigma_{Hjj}$ (fb) & 132 & 77 & 57.6 \\
\end{tabular}
\end{table}

The resulting $Hjj,\;H\to\tau\tau$ cross section is compared with the
irreducible $Zjj,\;Z\to\tau\tau$ backgrounds in the first row of
Table~\ref{table2}. Somewhat surprisingly, the EW $Zjj$ background reaches
$5\%$ of the QCD $Zjj$ background already at this level, while naively
one might expect suppression by a factor $(\alpha_{QED}/\alpha_s)^2
\approx 4\times 10^{-3}$. In the EW $Zjj$ background, $W$ exchange processes
can produce central $\tau$ pairs by $Z$ emission from the exchanged $W$ and
are therefore kinematically similar to the signal. This signal-like component 
remains after the forward jet tagging cuts, and, as we will see, will grow
in relative importance as the overall signal/background ratio is improved.

\begin{table}
\caption{Signal and background cross sections $B\sigma$ (fb) for
$m_H = 120$~GeV$
Hjj$ events in $pp$ collisions at $\protect\sqrt{s}=14$~TeV. Results are given 
after increasingly stringent cuts. The last column gives the ratio of the
signal to the background cross sections listed in the previous columns.
}
\label{table2}
\begin{tabular}{l|ccccc|c}
\phantom{generic} &
    $Hjj$ & QCD $Zjj$ & EW $Zjj$ & $Wj+jj$ & $b\bar{b}+jj$ & S/B \\
\hline
~forward tagging [Eqs.~(\ref{eq:basic},\ref{eq:taucen},\ref{eq:gap})] &
                      57.6 & 1670 & 90 &  &  &  \\
$+ \; \tau$ identification [Eq.~(\ref{eq:tauID})] &
        1.79 & 20.0 & 1.44 & 26.4 & 7.6 &  1/30  \\
$+ \; 110 < m_{\tau\tau} < 130 {\rm GeV}$ [Eq.~(\ref{eq:taumass2})] &
        1.18 & 0.95 & 0.07 & 1.77 & 0.6 & 1/3 \\
$+ \; m_{jj}>1$~TeV, $m_T(\ell,\sla p_T) < 30$~GeV & & & & & & \\
~[Eqs.~(\ref{eq:mjj},\ref{eq:mTlnu})] &
                      0.62 & 0.17 & 0.04 & 0.11 & 0.15 & 1.3/1 \\
$+ \; x_{\tau_l} < 0.75$, $x_{\tau_h} < 1.0$ [Eq.~(\ref{eq:x1x2})] &
                      0.49 & 0.14 & 0.03 & 0.02 & 0.05 &  2/1 \\
\end{tabular}
\end{table}

So far we have not considered $\tau$ decays. In order to get more realistic
rate estimates and to include the reducible backgrounds ($Wj+jj$ and
$b\bar{b}jj$, see Section~\ref{sec:four}) we need to study definite $\tau$
decay channels. We consider $\tau^+\tau^-$ decays with one $\tau$ decaying
leptonically ($e$ or $\mu$) and the other decaying hadronically in the
following, since previous studies have shown that dual leptonic decay is
more difficult to observe~\cite{Cavalli}. With a hadronic branching ratio
$B(\tau\to\nu+{\rm hadrons})=0.65$ and the overall hadronic $\tau$-decay
identification efficiency of Eq.~(\ref{eq:epstauh}), the selection of this
$\tau$-pair decay channel immediately reduces all $\tau^+\tau^-$ rates by a
factor
\ba
\epsilon B & = & 2 \epsilon_\tau(\tau\to\,\nu+{\rm hadrons}) \;
B(\tau\to\nu+{\rm hadrons})\; B(\tau\to\ell\nu_\ell\nu_\tau) \nonumber \\
  & = & 2\cdot 0.26\cdot 0.65\cdot 0.35 = 1/8.5 \, .
\ea
In addition, triggering the event via the isolated $\tau$-decay lepton and
identifying the hadronic $\tau$ decay as discussed in Ref.~\cite{Cavalli}
requires sizable transverse momenta for the observable $\tau$ decay products.
In the following we require
\bq
\label{eq:tauID}
p_{T_{\tau,lep}} > 20~{\rm GeV} \, , \qquad
p_{T_{\tau,had}} > 40~{\rm GeV} \, ,
\eq
where the second requirement is needed to use the results of
Cavalli {\it et al.} on hadronic $\tau$ identification.
These transverse momentum requirements
are quite severe and reduce the Higgs signal by another factor of $3.8$.
Resulting signal and background cross sections are given in the second row
of Table~\ref{table2}.

Crucial for further background reduction is the observation that the
$\tau$-pair invariant mass can be reconstructed from the observable $\tau$
decay products and the missing transverse momentum vector of the
event~\cite{tautaumass}.
Denoting by $x_{\tau_i}$ the fractions of the parent $\tau$ energy which
each observable decay particle carries, the transverse momentum vectors
are related by
\bq
\label{eq:taurecon}
\vec\sla{p_T} = ({1\over x_{\tau_l}} - 1) \; \vec p_{\ell} +
({1\over x_{\tau_h}} - 1) \; \vec p_h \; .
\eq
Here we neglect the $\tau$ mass and assume that the neutrinos from the
$\tau$ decays are collinear with the charged observables, a condition
which is satisfied to an excellent degree because of the high $\tau$
transverse momenta needed to satisfy Eq.~(\ref{eq:tauID}). As long as the
the decay products are not back-to-back, Eq.~(\ref{eq:taurecon}) gives two
conditions for $x_{\tau_i}$ and provides  the $\tau$ momenta as
$\vec p_{\ell}/x_{\tau_l}$ and $\vec p_h/x_{\tau_h}$, respectively.
This last condition is met in our case because the $H$ and $Z$ bosons
are typically produced with high $p_T$, on the order of 150~GeV for all
processes except the $b\bar{b}jj$ background (in which case the average
$p_T \approx 85$~GeV is still sufficient).

Mismeasured transverse momenta (smearing effects) can still lead to
unphysical solutions for the reconstructed $\tau$ momenta. In order to avoid
these, we impose a cut on the angle between the $\tau$ decay products
and require positivity of the calculated $x_{\tau_i}$:
\bq
\label{eq:taumass2}
\cos\theta_{\tau\tau} > -0.9 \, , \qquad  x_{\tau_{l,h}} > 0 \, .
\eq

The resulting $\tau$-pair invariant mass resolution is somewhat narrower
than the one found in Ref.~\cite{Cavalli}, the 1-$\sigma$ half-width for
the $H$ peak ranging from about 7 GeV for $m_H = 110$~GeV to about 10 GeV
for $m_H = 150$~GeV (see Fig.~\ref{fig:Mtautau} below).
This improved resolution is an effect of the higher
average $p_T$ of the underlying process: in our case, the two forward
tagging jets from weak boson scattering impart a higher $p_T$ on the
$H$ or $Z$ than is the case from QCD radiation in gluon fusion. The
smaller $\tau^+\tau^-$ opening angle then leads to a better $\tau$
momentum reconstruction via Eq.~(\ref{eq:taurecon}).
Given this $\tau$-pair mass resolution, we choose $\pm 10$~GeV mass bins
for analyzing the cross sections. Signal and background cross sections
in a 20~GeV mass bin centered at 120~GeV, after the reconstruction
conditions of Eq.~(\ref{eq:taumass2}), are listed in the third row of
Table~\ref{table2}. QCD and EW $Zjj$ backgrounds are reduced
by a factor of 20, while about 2/3 of the signal survives the mass
reconstruction cuts.

\begin{figure}[t]
\vspace*{0.5in}
\begin{picture}(0,0)(0,0)
\includegraphics{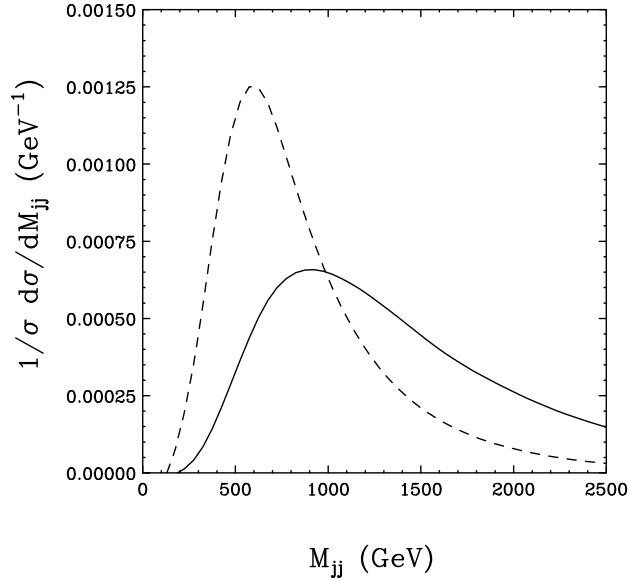}
\end{picture}
\vspace{7.0cm}
\caption{Invariant mass distribution of the two tagging jets for the $Hjj$
signal (solid line) and the QCD $Zjj$ background (dashed line), at the
level of forward tagging cuts and $\tau$ reconstruction,
Eqs.~(\ref{eq:basic}-\ref{eq:taumass2},\ref{eq:mTlnu}).
}
\vspace*{0.2in}
\label{fig:Mjj}
\end{figure}

Because the QCD backgrounds typically occur at small invariant masses, we
can further reduce them by imposing a cut on the invariant mass of the
tagging jets,
\bq
\label{eq:mjj}
m_{jj} > 1~{\rm TeV}.
\eq
Fig.~\ref{fig:Mjj} shows the tagging jets' invariant mass distribution for
the signal and QCD $Zjj$ background to illustrate the effect of the cut.

\section{Fake $\tau^{+}\tau^{-}$ events: reducible backgrounds}
\label{sec:four}

Reducible backgrounds to the $H\to\tau\tau$ signal, with subsequent leptonic
decay of one of the $\tau$'s, arise from any source of isolated, single
hard leptons.
As discussed in Section~\ref{sec:two}, we consider $Wj+jj$ events and heavy
quark production, in the form of $b\bar{b}jj$ events. Intrinsically, these
reducible backgrounds are enormous and overwhelm even the physics
backgrounds before $\tau$ identification and tight lepton isolation cuts
are made. Crucial for the reduction of these backgrounds to a manageable
level is the requirement of a narrow $\tau$-like jet, which leads to a
factor 400 suppression for the $Wj+jj$ background
(see Section~\ref{sec:twoD}).
The probability for a $b$-quark to fluctuate into a narrow $\tau$-like jet
is even smaller, below 0.0015, and another large reduction, by a
factor 100 (see Section~\ref{sec:twoE}), is expected from requiring the
$b$-decay lepton to be well isolated. An additional factor of two reduction
is achieved by requiring opposite charges for the isolated lepton and the
tau-like jet. The resulting background rates,
for charged leptons and $\tau$-like jets satisfying the transverse momentum
requirements of Eq.~(\ref{eq:tauID}), are listed in the second
row of Table~\ref{table2}.

Unlike the Higgs signal or the $Zjj$ backgrounds, the reducible backgrounds
show no resonance peaks in the $m_{\tau\tau}$ distribution. As a result,
another reduction by an order of magnitude is achieved when comparing rates
in a Higgs search bin of width 20~GeV (third row of Table~\ref{table2}).
Additional reductions are possible by making use of specific properties
of the reducible backgrounds. Analogous to the QCD $Zjj$ background, the
$Wj+jj$ and $b\bar{b}jj$ backgrounds are created at smaller parton center of
mass energies than the signal. As a result, the $m_{jj}>1$~TeV cut of
Eq.~(\ref{eq:mjj}) reduces both of them by roughly a factor of 4.

\begin{figure}[t]
\vspace*{0.5in}
\begin{picture}(0,0)(0,0)
\includegraphics{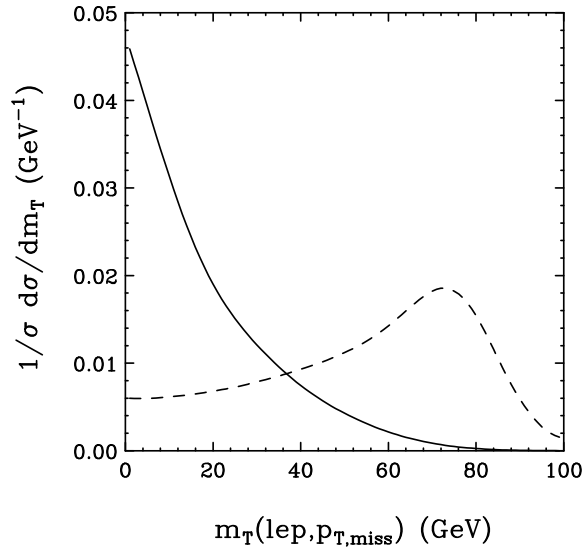}
\end{picture}
\vspace{6.5cm}
\caption{Transverse mass distribution of the $\ell$-$\sla p_T$ system for
the $Hjj$ signal (solid line) and the $Wj+jj$ reducible background (dashed
line), at the level
of far forward tagging cuts, $\tau$-reconstruction, and $m_{jj} > 1$~TeV
(Eqs.~\ref{eq:basic}-\ref{eq:mjj}).
}
\vspace*{0.2in}
\label{fig:mT}
\end{figure}

Further suppression of the $Wj+jj$ background can be achieved by taking
advantage of the Jacobian peak in the lepton-$\sla p_T$ transverse mass
distribution~\cite{Cavalli},
a feature which is otherwise used to measure the mass of
the $W$. We compare the $m_T$ distribution for the signal and the $Wj+jj$
background in Fig.~\ref{fig:mT}. A cut
\bq
\label{eq:mTlnu}
m_T(\ell,\sla p_T) < 30~{\rm GeV} \,
\eq
reduces the $Wj+jj$ background by a factor of 5 while reducing the signal
acceptance by only $15\%$. Similar to the signal, the other backgrounds
are affected very little by the transverse mass cut.

\begin{figure}[t]
\vspace*{0.5in}
\begin{picture}(0,0)(0,0)
\includegraphics{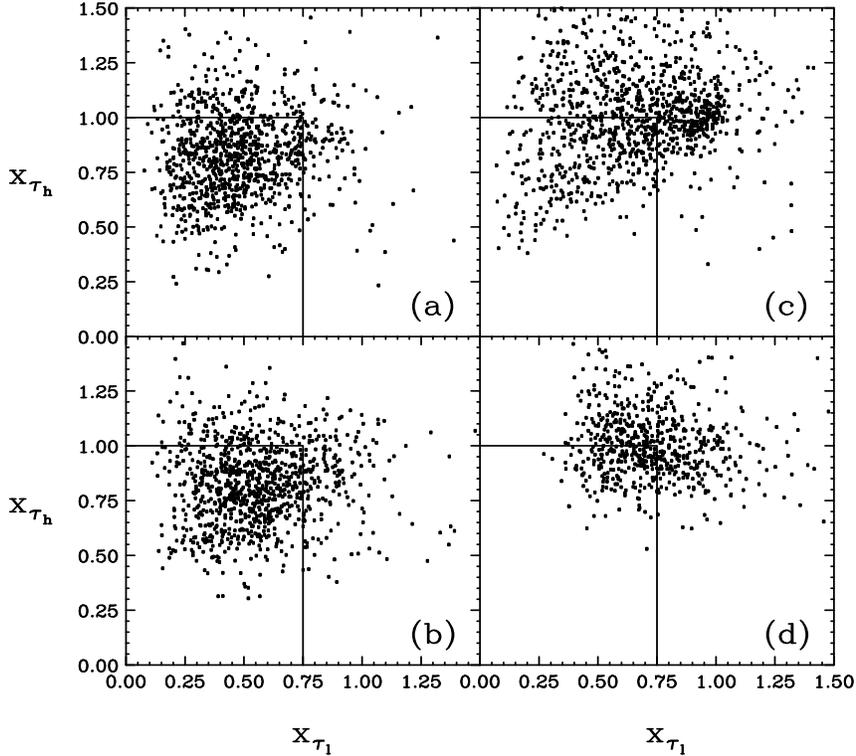}
\end{picture}
\vspace{8.5cm}
\caption{Scatter plots of $x_{\tau_l}$ vs. $x_{\tau_h}$ with the cuts of
Eqs.(\ref{eq:basic}-\ref{eq:gap},\ref{eq:tauID},
\ref{eq:taumass2}-\ref{eq:mTlnu}), for:
(a) the 120 GeV $Hjj$ signal; (b) the combined QCD and EW $Zjj$ irreducible
backgrounds; (c) the $Wj+jj$ and (d) the $b\bar{b}jj$ reducible backgrounds.
The number of points in each plot is arbitrary and corresponds to
significantly higher integrated luminosities than expected for the LHC.
The solid lines indicate the cuts of Eq.~(\ref{eq:x1x2}).
}
\vspace*{0.2in}
\label{fig:x1x2}
\end{figure}

At this level the S/B ratio is nearly 1/1, and we can study
additional event characteristics, such as the missing momentum.
In real $\tau$-pair events, the missing
momentum is a vector combination of neutrino momenta, which carry away
a significant fraction of the $\tau^+$ and $\tau^-$ energies.
In the reducible backgrounds it is purely from the leptonically
decaying parent particle, either the $W$ or one of the $b$'s. As such, we
should reconstruct $x_{\tau_h} = 1$ for the narrow, $\tau$-like jet, except
for smearing effects. The effect is clearly observable in the distribution
of events in the $x_{\tau_l}$--$x_{\tau_h}$ plane, which is shown
in Fig.~\ref{fig:x1x2}. The $x_{\tau_l}$ distribution of the leptonically
decaying $\tau$-candidate also is softer for real $\tau$'s than for the
reducible backgrounds, because the charged lepton shares the parent $\tau$
energy with two neutrinos. A cut
\bq
\label{eq:x1x2}
x_{\tau_l} < 0.75 \, , \qquad   x_{\tau_h} < 1 \, ,
\eq
proves very effective in suppressing the reducible backgrounds. For the
$Wj+jj$ background we find suppression by another factor of 4.5 and the
$b\bar{b}jj$ background is reduced by a factor of 3, while retaining
$80\%$ of the signal rate. One should note that these cuts are not
optimized, they are merely chosen to demonstrate the usefulness of the
$x_{\tau_l}$--$x_{\tau_h}$ distributions in restricting the otherwise
troublesome reducible backgrounds to a manageable level. Cross sections
including these cuts are given in the last row of Table~\ref{table2}.

In principle, the $x_\tau$ distributions contain information on $\tau$
polarization and $x_{\tau_l}$--$x_{\tau_h}$ correlations allow one to
distinguish between the decay of a spin-0 object, like the Higgs which
results in opposite $\tau^+$ and $\tau^-$ chiralities, and the decay of
the spin-1 $Z$ boson, with equal $\tau^\pm$ chiralities~\cite{taupolzn}.
Comparison of the two scatter plots in Fig.~(\ref{fig:x1x2}a) and
(\ref{fig:x1x2}b) shows, however, that the remaining correlations are very
weak. This may partially be due to the stringent transverse momentum cuts
(\ref{eq:tauID}) on the $\tau$ decay products which needed to be
imposed for background reduction.
In addition, the visible $\tau$ energy fractions in
$\tau\to\ell\bar{\nu_\ell}\nu_\tau$ and $\tau\to\rho\nu_\tau$ decays
are mediocre polarization analyzers only (measuring the splitting of the
$\rho$'s energy between its two decay pions would improve the situation for
the latter~\cite{HMZ}). A dedicated study is needed to decide whether
a $\tau$ polarization analysis is feasible at the LHC, but because of the
small rates implied by Table~\ref{table2} we do not pursue this issue here.

\section{Radiation patterns of minijets}\label{sec:five}

A further characteristic of EW vs. QCD scattering can be exploited,
namely the absence of color exchange between the two scattering quarks
in the $qq\to qqH$ signal process. $t$-channel color singlet
exchange in the EW case leads to soft gluon emission mainly in the very
forward and very backward directions, whereas QCD processes are dominated
by $t$-channel color octet exchange which results in soft gluon radiation
mainly in the central detector. It was hoped that resulting rapidity gaps
in signal events (large regions in pseudorapidity without observed hadrons)
could be used for background suppression~\cite{bjgap}. Unfortunately, in
$pp$ collisions of $\sqrt{s}=14$~TeV at the LHC, overlapping events in a
single bunch crossing will likely fill a rapidity gap even if it is present
at the level of a single $pp$ collision. Very low luminosity running
is not useful because of the small signal cross section.

The different
color structures of signal and background processes can be exploited even
at high luminosity, however, if one defines rapidity gaps in terms of
minijets ($p_{Tj} \approx$ 15-40~GeV) instead of soft
hadrons~\cite{bpz_minijet}. As has been shown for the analogous EW $Zjj$
process~\cite{RSZ_vnj}, with its very similar kinematics, minijet emission
in EW exchange occurs mainly in the very forward and very backward regions,
and even here is substantially softer than in the QCD $Zjj$
background. A veto on these central minijets will substantially improve the
signal-to-background ratio. Following the analysis of Ref.~\cite{RSZ_vnj}
we veto additional central jets in the region
\begin{mathletters}\label{eq:veto}
\begin{eqnarray}
p_{Tj}^{\rm veto} & > & p_{T,{\rm veto}}\;, \label{eq:ptveto} \\
\eta_{j,min}^{\rm tag} +0.7 & < & \eta_j^{\rm veto}
< \eta_{j,max}^{\rm tag} -0.7\; , \label{eq:etaveto}
\end{eqnarray}
\end{mathletters}
where $p_{T,\rm veto}$ may be chosen based on the capability of the detector.

Sizable background reduction via a minijet veto requires the lowering of the
$p_{T,\rm veto}$ threshold to a range where the probability for additional
parton emission becomes order unity. In a perturbative calculation the
resulting condition,
$\sigma(n+1\;{\rm jets})\approx \sigma(n\;{\rm jets})$, indicates
that one is leaving the validity range of fixed-order perturbation theory,
and it becomes difficult to provide reliable theoretical estimates of minijet 
emission rates. Gluon emission is governed by very different scales in signal 
as compared to background processes, due to their different color structures. 
Thus, a parton shower approach does not immediately give reliable answers
unless both color coherence and the choice of scale are implemented
correctly, corresponding to the answer given by a complete QCD calculation.

The necessary additional information on angular distributions and hardness
of additional radiation is available in the ``3 jet'' programs discussed in
Section~\ref{sec:two}. However, cross sections evaluated with these
minijet emission codes exceed the hard-process cross sections at moderate
transverse momenta of the additional jet already, namely at
$p_{T,\rm veto} \approx 40$~GeV for the QCD cases
and $\approx 10$~GeV for the EW cases. In order to extract meaningful
estimates, with $p_{T,\rm veto} \approx 15-20$~GeV, one needs
to regulate the $p_{Tj}\to 0$ singularities.  We use the
truncated shower approximation (TSA)~\cite{TSA} for this purpose,
which simulates the effects of soft multiple-gluon emission by replacing the
tree-level 3 jet differential cross section, $d\sigma_3^{\rm TL}$, with
\begin{equation}\label{eq:tsa}
d\sigma_3^{\rm TSA}=d\sigma_3^{\rm TL}
\left(1-e^{-p_{T3}^2/p_{TSA}^2}\right)\;.
\end{equation}
Here the parameter $p_{TSA}$ is chosen to correctly reproduce the
tree-level 2 jet cross section, $\sigma_2$, within the cuts of
Eqs.~(\ref{eq:basic},\ref{eq:tauID},\ref{eq:taumass2}-\ref{eq:mTlnu}), 
{\it i.e.} $p_{TSA}$ is fixed by the matching condition
\bq
\sigma_2 = \int_0^\infty {d\sigma_3^{\rm TSA}\over dp_{T3}} dp_{T3}\; .
\eq
We find $p_{\rm TSA} = 6.7$~GeV for the $Hjj$ signal,
$p_{\rm TSA} = 12.1$~GeV for the EW $Zjj$ background, and
$p_{\rm TSA} = 60$~GeV for the QCD $Zjj$ and $Wj+jj$ backgrounds.
The much larger value for the QCD
processes again reflects the higher intrinsic momentum scale governing
soft-gluon emission in the QCD backgrounds. This difference is enhanced even
more by requiring larger dijet invariant masses for the two tagging
jets~\cite{RSZ_vnj}.

Using $d\sigma_3^{\rm TSA}$ as a model for additional jet activity,
we can study the efficiency of vetoing central soft jet emission. The
survival probability for signal and background processes is found by
rejecting events with a minijet of $p_{Tj}^{\rm veto} > p_{T,veto}$ in the
gap region (\ref{eq:etaveto}), and by dividing the resultant, regulated
cross section by the inclusive ($2j$) cross section. Results are summarized in 
Table~\ref{table3}. In order to determine these numbers, we must first
select two of the three final state partons as tagging jets, for which
several methods exist.
In the first, dubbed ``$p_T$-method'', we choose the two
jets with the highest $p_T$'s, because the quark jets of the signal are
typically much harder than gluon jets from additional soft radiation.
Two other choices, the ``R-method'' and ``$\eta$-method'', select the
two jets closest to the reconstructed Higgs boson in $\triangle{R}$
and $\triangle{\eta}$, respectively, because additional radiation
in the signal is mainly expected in the very far forward regions, at larger
separations from the Higgs boson than the quark jets.
The R- and $\eta$-methods give slightly higher signal significances,
but are still consistent with the $p_T$-method. Results in Table~\ref{table3}
were derived with the $\eta$-method.

The minijet veto
reduces the signal by about $30\%$, but eliminates typically $85\%$ of the
QCD backgrounds. The EW $Zjj$ background is reduced by about $50\%$,
reflecting a radiation pattern for the $t$-channel $W$-exchange graphs
which is similar to the signal process, but also indicates the presence of
additional bremsstrahlung processes which allow radiation back into the
central region. In addition, the exchanged transverse $W$'s in the
EW $Zjj$ case result in higher-$p_T$ quark jets, on average, than the
longitudinal $W$'s that are exchanged in the $Hjj$ signal. This is also
reflected in the slightly higher value for $p_{\rm TSA}$ in the EW $Zjj$
case as compared to the $Hjj$ signal.

\begin{table}
\caption{Survival probabilities for the signal and backgrounds, using the
$\eta$-method for selecting the tagging jets, and for $p_{T,veto} = 20$~GeV.
The second row gives the number of events expected for
30~${\rm fb}^{-1}$ of integrated luminosity, after application of all cuts,
Eqs.(\ref{eq:basic}-\ref{eq:gap},\ref{eq:tauID},
\ref{eq:taumass2}-\ref{eq:x1x2}), and for
$m_H = 120$~GeV and $110 < m_{\tau\tau} < 130$~GeV. Survival probabilities
for the $b\bar{b}jj$ background are assumed to be the same as for the $Wj+jj$ 
background. As a measure of the Poisson probability of the background to
fluctuate up to the signal level, the last column gives $\sigma_{Gauss}$,
the number of Gaussian equivalent standard deviations.
}
\label{table3}
\begin{tabular}{l|ccccc|c}
\phantom{generic} & $Hjj$ & QCD $Zjj$ & EW $Zjj$
                  & $Wj+jj$ & $\,\;b\bar{b}jj\;\;$ & $\sigma_{Gauss}$ \\
\hline
$P_{surv}$     & 0.71 & 0.14 & 0.48 & 0.15 & 0.15 &  \\
no. events     & 10.4 & 0.61 & 0.46 & 0.11 & 0.24 & 5.2 \\
\end{tabular}
\end{table}

Table~\ref{table4} applies the survival probabilities found for the
$\eta$-method to the cross sections after final cuts, for Higgs boson masses
ranging from 110 to 150~GeV. A constant size of the mass bins of
20~GeV is kept for simplicity. In the actual experiment, the mass window
will need to be optimized depending on the predicted width of the signal and
background distributions, and may have to be asymmetric for low values of
$m_H$. Our table merely shows how observing a light Higgs boson is
quite feasible, even in the mass window close to the smeared $Z$ peak. As
$m_H$ approaches 150~GeV, however, the $H \to \tau\tau$ branching ratio drops 
rapidly in the SM and the signal gets low for integrated luminosities
of order 30~fb$^{-1}$. It should be
noted that with higher luminosity, this channel is still very effective to
make a direct measurement of the $H\tau\tau$ coupling.

\begin{table}
\caption{Number of expected events for the signal and backgrounds,
for 30~${\rm fb}^{-1}$ integrated luminosity and cuts as in Table~\ref{table3},
but for a range of Higgs boson masses. Mass bins of $\pm 10$~GeV
around a given central value are assumed.
}
\label{table4}
\begin{tabular}{c|ccccc|c}
$m_H$(GeV) & $Hjj$ & QCD $Zjj$
       & EW $Zjj$ & $Wj+jj$ & $\;\;b\bar{b}jj\;\;$ & $\sigma_{Gauss}$ \\
\hline
110 & 11.1 & 2.1  & 1.4  & 0.1 & 0.3 & 4.1 \\
120 & 10.4 & 0.6  & 0.5  & 0.1 & 0.2 & 5.2 \\
130 &  8.6 & 0.3  & 0.3  & 0.1 & 0.2 & 5.0 \\
140 &  5.8 & 0.2  & 0.2  & 0.1 & 0.2 & 3.9 \\
150 &  3.0 & 0.1  & 0.2  & 0.1 & 0.2 & 2.3 \\
\end{tabular}
\end{table}

\section{Discussion}\label{sec:six}

\begin{figure}[htb]
\vspace*{0.5in}
\begin{picture}(0,0)(0,0)
\includegraphics{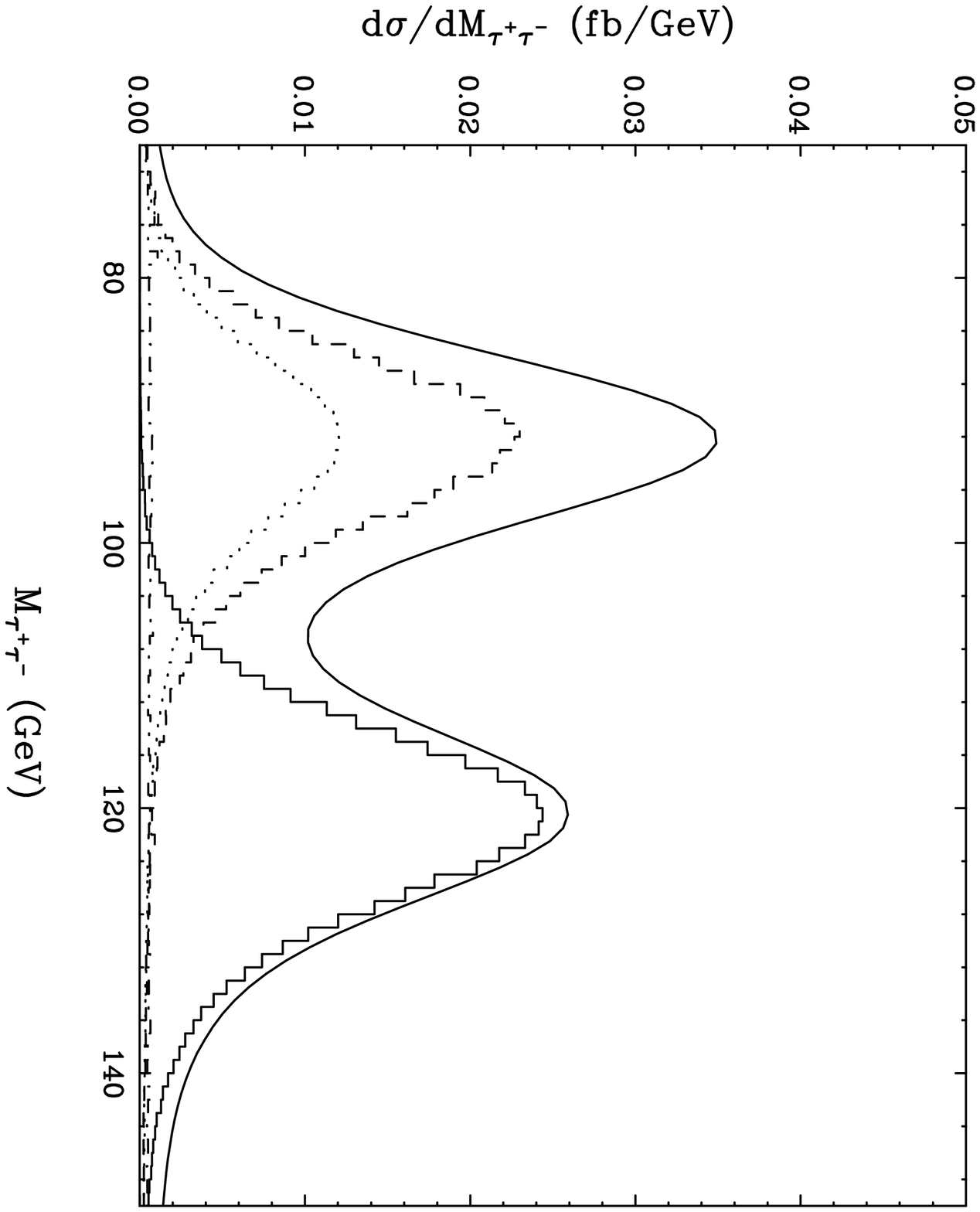}
\end{picture}
\vspace{9.0 cm}
\caption{Reconstructed $\tau$ pair invariant mass distribution for the
signal and backgrounds
after the cuts of Eqs.~(\ref{eq:basic}-\ref{eq:x1x2}) and
multiplication of the Monte Carlo results by the expected survival
probabilities.
The solid line represents the sum of the signal and all backgrounds.
Individual components are shown as histograms:
the $Hjj$ signal (solid), the irreducible QCD $Zjj$
background (dashed), the irreducible EW $Zjj$ background (dotted), and the
combined $Wj+jj$ and $b\bar{b}jj$ reducible backgrounds (dash-dotted).
}
\label{fig:Mtautau}
\vspace*{0.2in}
\end{figure}

The results summarized in Table~\ref{table4} show that it is possible to
isolate a virtually background free $qq\to qqH,\;H\to\tau\tau$ signal
at the LHC, with sufficiently large counting rate to obtain a $5\sigma$
signal with a mere 30~fb$^{-1}$ of data. The expected purity of the signal
is demonstrated in Fig.~\ref{fig:Mtautau}, where the reconstructed
$\tau\tau$ invariant mass distribution for a SM Higgs boson of mass 120~GeV
is shown, together with the various backgrounds, after application of all
cuts discussed in the previous Section.
This purity is made possible because the weak boson fusion process, together
with the $H\to\tau^+\tau^-\to\ell^\pm {\rm hadrons}^\mp \sla{p_T}$ decay,
provides a complex signal, with a multitude of characteristics which
distinguish it from the various backgrounds.

\begin{figure}[t]
\vspace*{0.5in}
\begin{picture}(0,0)(0,0)
\includegraphics{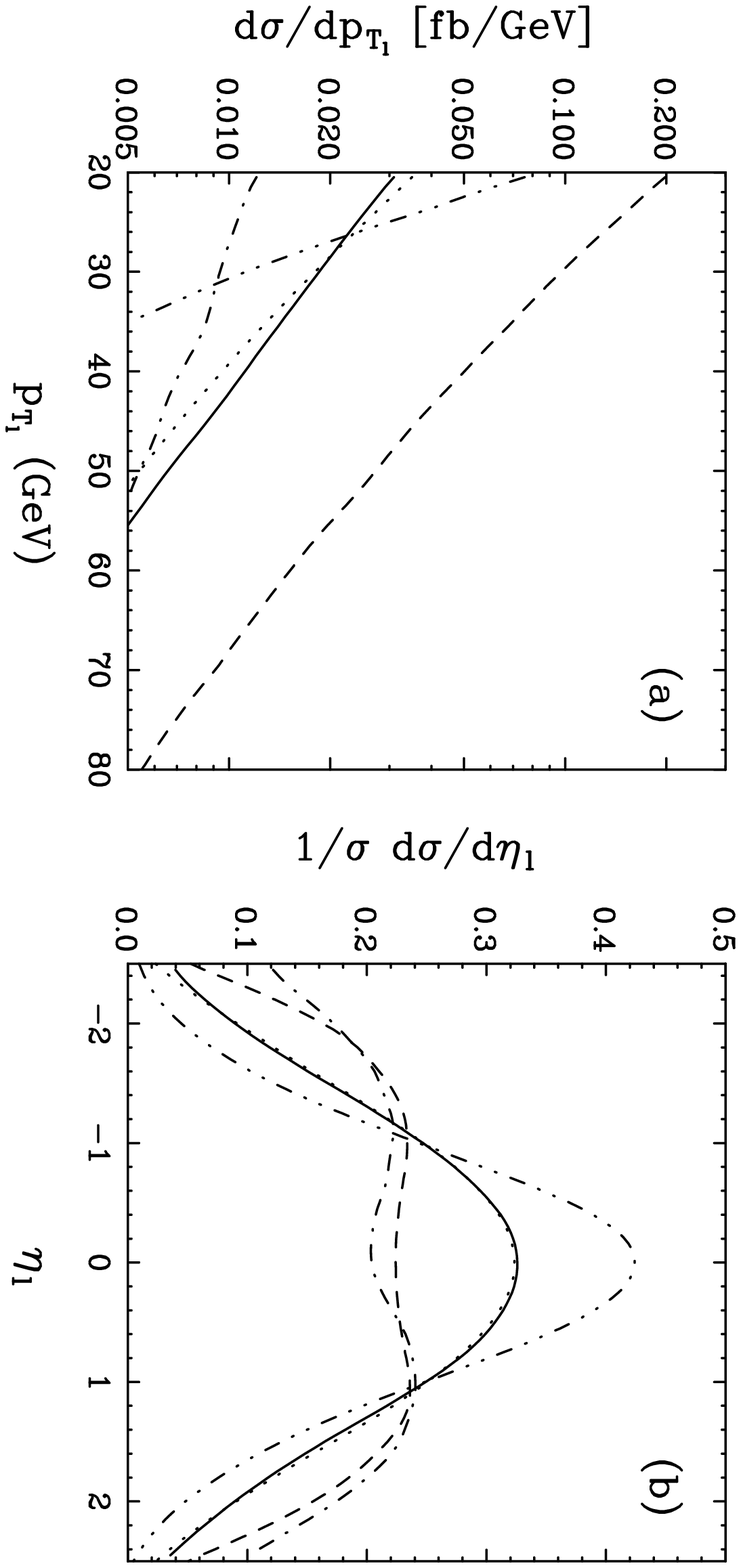}
\end{picture}
\vspace{6.5cm}
\caption{(a) Transverse momentum and (b) pseudorapidity distributions of
the charged ``$\tau$'' decay lepton after the cuts of 
Eqs.~(\ref{eq:basic}-\ref{eq:x1x2}), for the $m_H = 120$~GeV signal (solid
line), and backgrounds: QCD $Zjj$ production (dashed line), EW $Zjj$ events
(dotted line), $Wj+jj$ events (dot-dashed line), and $b\bar{b}jj$ production
(dash-double dotted line).
}
\label{fig:lepton}
\vspace*{0.2in}
\end{figure}

The basic feature of the $qq\to qqH$ signal is the presence of two forward
tagging jets inside the acceptance of the LHC detectors, of sizable $p_T$,
and of dijet invariant mass in the TeV range. Typical QCD backgrounds,
with isolated charged leptons and two hard jets, are much softer.
In addition, the QCD backgrounds are dominated by $Z$ or $W$ bremsstrahlung
off forward scattered quarks, which gives typically higher-rapidity
charged leptons (see Fig.~\ref{fig:lepton}). In contrast,
the EW processes give rise to quite central leptons, and this includes not
only the Higgs signal but also EW $Zjj$ production, which also proceeds via
weak boson fusion. It is this similarity that prevents one from ignoring
EW $Zjj$ processes, which a priori are smaller by two orders of
magnitude in total cross section, but after final cuts remain the same
size as their QCD counterparts.

We advocate taking advantage of an
additional fundamental characteristic of QCD and EW processes. Color-singlet
exchange in the $t$-channel, as encountered in Higgs boson production by weak 
boson fusion (and in the EW $Zjj$ background), leads to additional soft jet
activity
which differs strikingly from that expected for the QCD backgrounds in both
geometry and hardness: gluon radiation in QCD processes is typically both
more central and harder than in WBF processes. We exploit this
radiation, via a veto on events with central minijets, and expect a
typical $85\%$
reduction in QCD backgrounds, but only about a $30\%$ loss of the signal.

The properties mentioned so far are generic in the search for weak boson
fusion events. Additional cuts are specific to the $H\to\tau\tau$ channel,
with one $\tau$ decaying leptonically and the other one decaying hadronically.
Crucial are charged lepton isolation and efficient identification of the
hadronically decaying $\tau$, which are needed for the suppression of heavy
quark backgrounds and non-$\tau$ hadronic jets. This part of the analysis
we have adapted from Ref.~\cite{Cavalli}, which, however, was performed
for $A,H\to\tau\tau$ events from gluon fusion, i.e. without requiring two
additional forward tagging jets. A more detailed assessment of lepton isolation
and hadronic $\tau$ identification in the present context is beyond the scope
of the present work and should be performed with a full detector simulation.

The elimination of the $Wj+jj$ reducible background depends highly
upon the Jacobian peak in the transverse mass distribution
of the $W$ decay products. The other backgrounds and the Higgs signal
typically produce rather small values of $m_T(\ell,\sla{p_T})$, below
30~GeV, and thus well below the peak in $m_T(W)$.

\begin{figure}[t]
\vspace*{0.5in}
\begin{picture}(0,0)(0,0)
\includegraphics{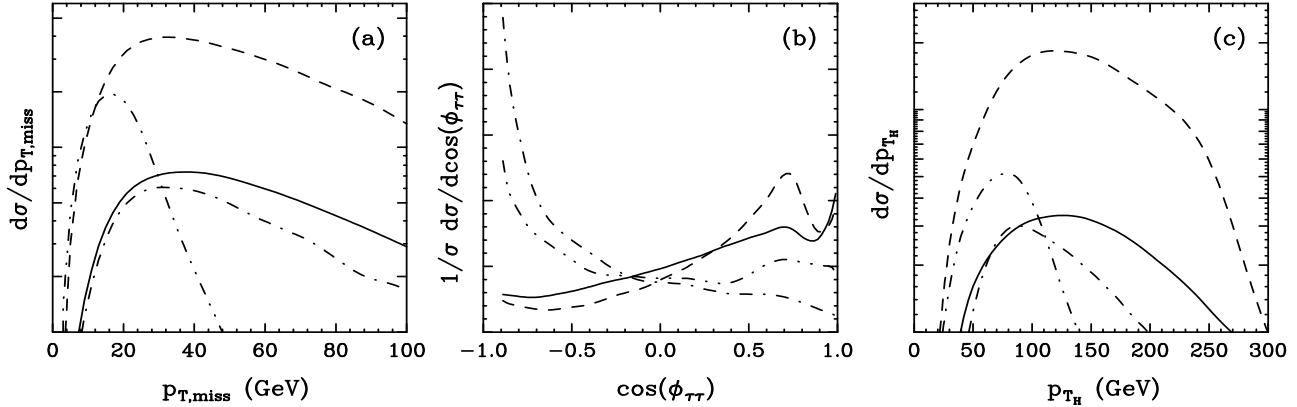}
\end{picture}
\vspace{5.0cm}
\caption{
Shape comparison of various distributions for the Higgs signal (solid 
line) and the backgrounds: QCD $Zjj$ production (dashed line),
$Wj+jj$ events (dot-dashed line), and $b\bar{b}jj$ production
(dash-double dotted line). Shown are the (a) $\sla p_T$, (b)
cos($\phi_{\tau\tau}$) and (c) transverse momentum distribution of the 
reconstructed $\tau\tau$ system, after the cuts of 
Eqs.~(\ref{eq:basic}-\ref{eq:x1x2}).
}
\label{fig:taudata}
\vspace*{0.2in}
\end{figure}

Another distinguishing feature of real $\tau$ decays are the reconstructed
momentum fractions $x_{\tau_l}$ and $x_{\tau_h}$ of the charged decay lepton
and of the decay hadrons. Misidentified ``$\tau$'s'' tend to produce
unphysically large values for these momentum fractions and can thereby be
eliminated to a substantial degree (see Fig.~\ref{fig:x1x2}).
The reconstruction of these $\tau$ momentum fractions is possible since
the $\tau^+\tau^-$ pairs are typically being produced with sizable
transverse momenta (see Fig.~\ref{fig:taudata}c)). As a result back-to-back
$\tau^+\tau^-$ decay products are rare (see Fig.~\ref{fig:taudata}b)) and
this in turns allows the mass reconstruction of the $\tau$-pair, which
is crucial for the suppression of the main physics background, $Z\to\tau\tau$.

We have not made full use of the differences between the Higgs signal and
the various backgrounds in some of these distributions. Additional examples
are shown in Figures \ref{fig:lepton} and \ref{fig:taudata}.
Fig.~\ref{fig:lepton} shows the $p_{T\ell}$ and $\eta_\ell$ distributions for 
the observable charged lepton, which will form an important part of the
event trigger. As a result of the lepton isolation cut, the
$p_{T\ell}$ falloff
is considerably steeper for the $b\bar{b}jj$ background than for the signal
and the other backgrounds. Not much leeway is present in applying more
stringent cuts, however, without losing a substantial fraction of the signal. 
One can also take advantage of the $\eta_{\ell}$ distribution for
the QCD $Zjj$ background, which, at the final level of cuts, remains
important in particular for small values of the Higgs boson mass.

In addition to the lepton $p_T$, we may use the missing transverse momentum
of the event, $\sla p_T$, Fig.~\ref{fig:taudata}~(a), which is
exceptionally small for the $b\bar{b}jj$ background. In combination with a
more stringent cut on the $\tau$ pair opening angle, cos($\phi_{\tau\tau}$),
shown in Fig.~\ref{fig:taudata}~(b) (where an even more striking distinction
between the physics and the reducible processes is found),
both the $Wj+jj$ and $b\bar{b}jj$
backgrounds can be reduced even below the level discussed in
Section~\ref{sec:five}. Such a strategy, however, may not increase the
statistical significance of the signal. In fact we find that slightly looser
cuts, for example on the dijet invariant mass, $m_{jj}$, can somewhat increase
the significance of the signal while reducing the signal-to-background
ratio. These points demonstrate that we have not yet optimized
the search strategy for $H\to\tau\tau$ decays. This might be possible
by combining the information from all the distributions mentioned above in a
neural-net analysis.
It is premature at this stage, however, to perform such an analysis since
the issues of $\tau$-identification or of suppression of heavy quark decays
in a realistic detector need to be addressed simultaneously, for the specific
processes considered here.

Beyond the possibility of discovering the Higgs boson in the $H\to\tau\tau$
mode, or confirmation of its existence, the independent measurement of the
$H\tau\tau$ coupling will be another important reason to strive for
observation of $H\to\tau\tau$ decays at the LHC. For such a measurement,
via the analysis outlined in this paper, $\tau$-identification efficiencies,
minijet veto probabilities etc. must be precisely known. For calibration
purposes, the presence of the $Z\to\tau\tau$ peak in Fig.~\ref{fig:Mtautau}
will be of enormous benefit. The production rates of the QCD and EW $Zjj$
events can be reliably predicted and, thus, the observation of the
$Z\to\tau\tau$ peak allows for a direct experimental assessment of the
needed efficiencies, in a kinematic configuration which is very similar
to the Higgs signal.

Observation of the $H\to\tau\tau$ decay mode at the LHC, for the SM Higgs, and
for modest integrated luminosities appears to be a real possibility.
What is needed is that the Higgs boson lies in the mass range between
present LEP limits and about 150 GeV, where its $\tau\tau$ branching
fraction is sizable. In models beyond the SM prospects may be even better.
Weak boson fusion at the LHC will be an exciting process to study at the
LHC, for a weakly coupled Higgs sector just as much as for strong interactions
in the symmetry breaking sector of electroweak interactions.

%
%

\acknowledgements
We would like to thank Tao Han for useful discussions and use of his
programs for $b$ decay simulations.
This research was supported in part by the University of Wisconsin Research
Committee with funds granted by the Wisconsin Alumni Research Foundation and
in part by the U.~S.~Department of Energy under Contract
No.~DE-FG02-95ER40896.
The work of K.H. is supported in part by the JSPS-NSH Joint Research Project,
and in part by Grant-in-Aid for Scientific Research from the Ministry of
Education, Science and Culture of Japan.
D.R. was supported in part by the NSF Summer Institute in Japan program.

%
%

\end{document}